\documentclass[useAMS,usenatbib]{mnras}

\usepackage{epsfig}
\usepackage{aas_macros}
\usepackage{natbib}
\usepackage{times}
\usepackage{graphicx, bm, amssymb, xcolor}
\usepackage{verbatim}
\usepackage[all]{hypcap}
\usepackage{xspace}
\usepackage{multirow}
\usepackage{amsmath}
\usepackage{epstopdf}
\usepackage{pdflscape}
\usepackage{xspace}

\newcommand{\msun}{\mbox{M$_\odot$}}

\newcommand{\gyr}{\mbox{${\rm Gyr}$}}

\newcommand{\kpc}{\mbox{${\rm kpc}$}}

\newcommand{\feh}{\mbox{$[{\rm Fe}/{\rm H}]$}}
\newcommand{\be}{\begin{equation}}
\newcommand{\ee}{\end{equation}}
\newcommand{\bea}{\begin{eqnarray}}
\newcommand{\eea}{\end{eqnarray}}

\newcommand{\emosaics}{{\sc E-MOSAICS}\xspace}
\newcommand{\mosaics}{{\sc MOSAICS}\xspace}
\newcommand{\eagle}{{\sc EAGLE}\xspace}

\usepackage{etoolbox}
\makeatletter
\patchcmd\@combinedblfloats{\box\@outputbox}{\unvbox\@outputbox}{}{%
    \errmessage{\noexpand\@combinedblfloats could not be patched}%
}%
\makeatother

\title{The mass fraction of halo stars contributed by the disruption of globular clusters in the E-MOSAICS simulations}
\author{Marta~Reina-Campos$^{1}$\thanks{reina.campos@uni-heidelberg.de}, Meghan~E.~Hughes$^{2,3}$, J.~M.~Diederik~Kruijssen$^{1}$, \newauthor
 Joel~L.~Pfeffer$^{2}$, Nate~Bastian$^{2}$, Robert~A.~Crain$^{2}$, Andreas~Koch$^{1}$ and Eva~K.~Grebel$^{1}$\\
$^{1}$Astronomisches Rechen-Institut, Zentrum f\"{u}r Astronomie der Universit\"{a}t Heidelberg, M\"{o}nchhofstra\ss e 12-14, 69120 Heidelberg, Germany\\
$^{2}$Astrophysics Research Institute, Liverpool John Moores University, 146 Brownlow Hill, Liverpool L3 5RF, UK\\
$^{3}$European Southern Observatory, Karl-Schwarzschild-Stra\ss e 2, 85748, Garching, Germany}
\begin{document}

\date{}

\pagerange{\pageref{firstpage}--\pageref{lastpage}} \pubyear{2019}

\maketitle

\label{firstpage}

\begin{abstract}
Globular clusters (GCs) have been posited, alongside dwarf galaxies, as significant contributors to the field stellar population of the Galactic halo. In order to quantify their contribution, we examine the fraction of halo stars formed in stellar clusters in the suite of 25 present-day Milky Way-mass cosmological zoom simulations from the \emosaics project. We find that a median of $2.3$ and $0.3$ per cent of the mass in halo field stars formed in clusters and GCs, defined as clusters more massive than $5\times 10^3$ and $10^5~\msun$, respectively, with the $25$--$75$th percentiles spanning $1.9$--$3.0$ and $0.2$--$0.5$ per cent being caused by differences in the assembly histories of the host galaxies. Under the extreme assumption that no stellar cluster survives to the present day, the mass fractions increase to a median of $5.9$ and $1.8$ per cent. These small fractions indicate that the disruption of GCs plays a sub-dominant role in the build-up of the stellar halo. We also determine the contributed halo mass fraction that would present signatures of light-element abundance variations considered to be unique to GCs, and find that clusters and GCs would contribute a median of $1.1$ and $0.2$ per cent, respectively. We estimate the contributed fraction of GC stars to the Milky Way halo, based on recent surveys, and find upper limits of $2$--$5$ per cent (significantly lower than previous estimates), suggesting that models other than those invoking strong mass loss are required to describe the formation of chemically enriched stellar populations in GCs.
\end{abstract}


\section{Introduction} \label{sec:intro}
Understanding the formation and evolution of massive, compact stellar clusters, also known as globular clusters (GCs), allow the reconstruction of the assembly history of their host galaxies \citep[e.g.][]{forbes18,myeong18d,kruijssen19b,massari19}. Over the past decade, several studies have used the presence of light-element abundance variations in GCs \citep[characterized by a depletion in C, O, and Mg and an enhancement in N, Na, Al; e.g.][]{carretta09a,piotto15}, along with a chemical-tagging technique, to identify stars in our Galaxy as candidates that may have formed in stripped or dissolved GCs. This technique has been applied to the inner Galaxy \citep{schiavon17}, as well as to halo field stars to reconstruct the build-up of our Galaxy \citep[e.g.][]{martell10,martell11,carollo13,martell16,koch19}. These latter studies find that ${\sim}1.4$--$2.6$ per cent of halo field stars exhibit light-element abundance patterns resembling those of GCs. This suggests that, if GCs are the unique formation sites of chemically distinct stellar populations\footnote{We use the terms `chemically-distinct' or `enriched' stars interchangeably to refer to stars exhibiting the light-element abundance patterns resembling those of GCs, and we will refer to the stellar population without these chemical features as `unenriched' or `primordial'.}, an upper limit of $\sim 11$--$47$ per cent of halo stars have originated in GCs, with the exact number depending on the details of the GC formation and evolution model, as well as the fraction of enriched-to-unenriched stars considered \citep[][]{carretta16,martell16,koch19}. 

These fractions would imply that a considerable fraction of the stellar halo originated in disrupted or surviving GCs. However, recent observational studies suggest a different scenario for the formation of the Galactic stellar halo. By comparing the high blue straggler-to-blue horizontal branch stellar ratio in the stellar halo to the low ratios observed in GCs, \citet{deason15} argue that the different population ratios favour a scenario in which the Galactic stellar halo has been built up by a few, relatively massive dwarf galaxies. In addition, a large number of studies using data from the \textit{Gaia} mission suggest that the a large fraction of the inner Galactic stellar halo was contributed by a single satellite of mass $\sim 10^9~\msun$ that was accreted $\sim 9$--$10~\gyr$ ago (\textit{Gaia}-Enceladus/Sausage; e.g. \citealt{belokurov18,helmi18}; also see \citealt{kruijssen19b}). Recently, \citet{conroy19} find that the bulk of the stellar halo splits in discrete features in the orbital-chemical space, indicating that the majority of halo stars have assembled from tidally-disrupted dwarf galaxies.

In a companion paper, we look at the mass contribution of GCs to the bulge in the suite of 25 present-day Milky Way-mass galaxies from the \emosaics simulations \citep{pfeffer18,kruijssen19a}. We find that the disruption of GCs contributes between $0.3$--$14$ per cent of the bulge mass, in agreement with recent observational estimates \citep{hughes19b}. With the aim of determining whether the disruption of GCs plays a prominent role in the build-up of stellar haloes, in this work we quantify the total mass contribution of clusters and GCs, as well as that of chemically-enriched stars, to the stellar halo in the 25 Milky Way-mass galaxies from the \emosaics suite, which we then compare with results obtained from recent observational studies.

\section{Summary of the E-MOSAICS simulations}

In order to determine the contribution of the dynamically disrupted mass from clusters and GCs to the build-up of the stellar halo, we use the 25 cosmological zoom-in simulations of present-day Milky Way-mass galaxies that are part of the \emosaics suite. The MOdelling Star cluster population Assembly In Cosmological Simulations (\mosaics, \citealt{kruijssen11,pfeffer18}) within \eagle (Evolution and Assembly of GaLaxies and their Environments, \citealt{schaye15}, \citealt{crain15}) project combines a sub-grid description of bound stellar cluster formation and evolution with a state-of-the-art galaxy formation model within the $\Lambda$CDM cosmogony. This enables a self-consistent study of the formation and co-evolution of galaxies and their stellar cluster populations, in which GCs emerge from the cluster population after a Hubble time of evolution. For a detailed description of the physical models adopted in \emosaics and details of the simulations, we refer the reader to \citet{pfeffer18} and \citet{kruijssen19a}. Here we briefly summarize the most relevant prescriptions used.

Our description of cluster formation and evolution is as follows. Whenever a gas particle is converted to a stellar particle ($\sim 2.25\times10^5~\msun$), a cluster population forms within the stellar particle in a sub-grid fashion. The properties of the cluster population are governed by the fraction of stellar mass forming in bound clusters (i.e.~the cluster formation efficiency, \citealt{bastian08}) and the shape of the initial cluster mass function. The cluster formation efficiency is determined using the model of \citet{kruijssen12d}, which predicts a strong correlation with gas pressure that is also observed in nearby extragalactic systems \citep{adamo15b,johnson16}. The initial cluster mass function is assumed to be a Schechter function with an environmentally-dependent upper mass scale. This truncation mass is also predicted to increase with gas pressure \citep{reina-campos17}, and it is found to reproduce observations of young massive clusters in the local Universe \citep{reina-campos17,messa18,trujillogomez19}. Once formed, the clusters are evolved alongside their host galaxies in a cosmological context. The stellar clusters lose mass due to stellar evolution \citep{wiersma09b}, tidal shocks, two-body relaxation \citep{kruijssen11} and dynamical friction \citep{pfeffer18}, the latter being necessarily applied in post-processing. Such a description for cluster formation and evolution has been found to reproduce a wide variety of observed cluster populations (\citealt{pfeffer18,kruijssen19a,usher18,pfeffer19b}), as well as to predict links between the cluster population and its host galaxy \citep{kruijssen19a,hughes19,pfeffer19a,reina-campos19}.

\section{Mass fraction of halo stars formed in GCs}\label{sec:frac-emosaics}

In order to define the stellar halo of the central galaxy in our zoom-in simulations, we follow the same criteria described by \citet{zolotov09} (see their section~2.1). Using the present-day information in our simulations, we first determine the angular momentum in the $z$-direction (i.e. perpendicular to the disk), $J_{\rm z}$, of all stellar particles, and discard those that belong to the thin and thick disks, $J_{\rm z}/J_{\rm circ} \geq 0.5$ \citep{sales12}, i.e.~with angular momentum resembling that of a corrotating circular orbit with similar orbital energy, $J_{\rm circ}$. Once we select all stellar particles belonging to the spheroid ($J_{\rm z}/J_{\rm circ} < 0.5$ ), we distinguish between bulge and halo stars by considering a distance cut based on the half-mass stellar radius of each galaxy, $R_{\rm 1/2, *}$. Stars lying farther away than this radius and within $50~\kpc$, $R_{\rm 1/2, *}<r<50~\kpc$\footnote{This outer limit is chosen in order to facilitate the comparison with observational studies \citep[e.g.][]{koch19}.}, are considered to belong to the halo and we determine its mass from the halo field stellar population. According to this definition, we measure stellar halo masses of a median $\sim 3.4\times 10^9~\msun$ among our suite of galaxies, which decreases by a factor $\sim 6$ when restricted to a metallicity range typically used in chemical-tagging studies ($\feh\in[-1.8, -1.3]$, e.g.~\citealt{koch19}). These masses are lower limits, as the \eagle model is known to underpredict the peak of the ratio of stellar mass to halo mass of central galaxies \citep[][]{crain15}.

With the aim of comparing our results with those obtained through the chemical-tagging technique, we define our cluster and GC populations to resemble those in which chemically-distinct stellar populations have been observed. Hence, we define stellar clusters to be more massive than $m_{\rm cl}^{\rm init}\geq5\times10^3~\msun$ at birth\footnote{To reduce memory requirements, in \emosaics we consider that stellar clusters less massive will experience short disruption timescales (shorter than $1~\gyr$) and can be safely discarded at formation.}, older than $2~\gyr$ \citep{martocchia18}, more metal-rich than $[\rm Fe/H]>-3$~dex and part of the halo as described above. In addition to that, we restrict the GC population to be more massive than $m_{\rm cl}^{\rm init}\geq10^5~\msun$ at birth \citep{kruijssen15b}. 

Our cluster and GC populations are affected by dynamical friction, which we apply in post-processing. We assume that ex-situ objects disrupted by dynamical friction in their host dwarf galaxies contribute to the build-up of the stellar halo of the central galaxy when their host galaxy is accreted. On the contrary, in-situ objects that disrupt due to dynamical friction are assumed to sink into the center of the central galaxy, and their disrupted mass does not contribute to the build-up of the halo.

\begin{figure*}
\centering
\includegraphics[width=\hsize,keepaspectratio]{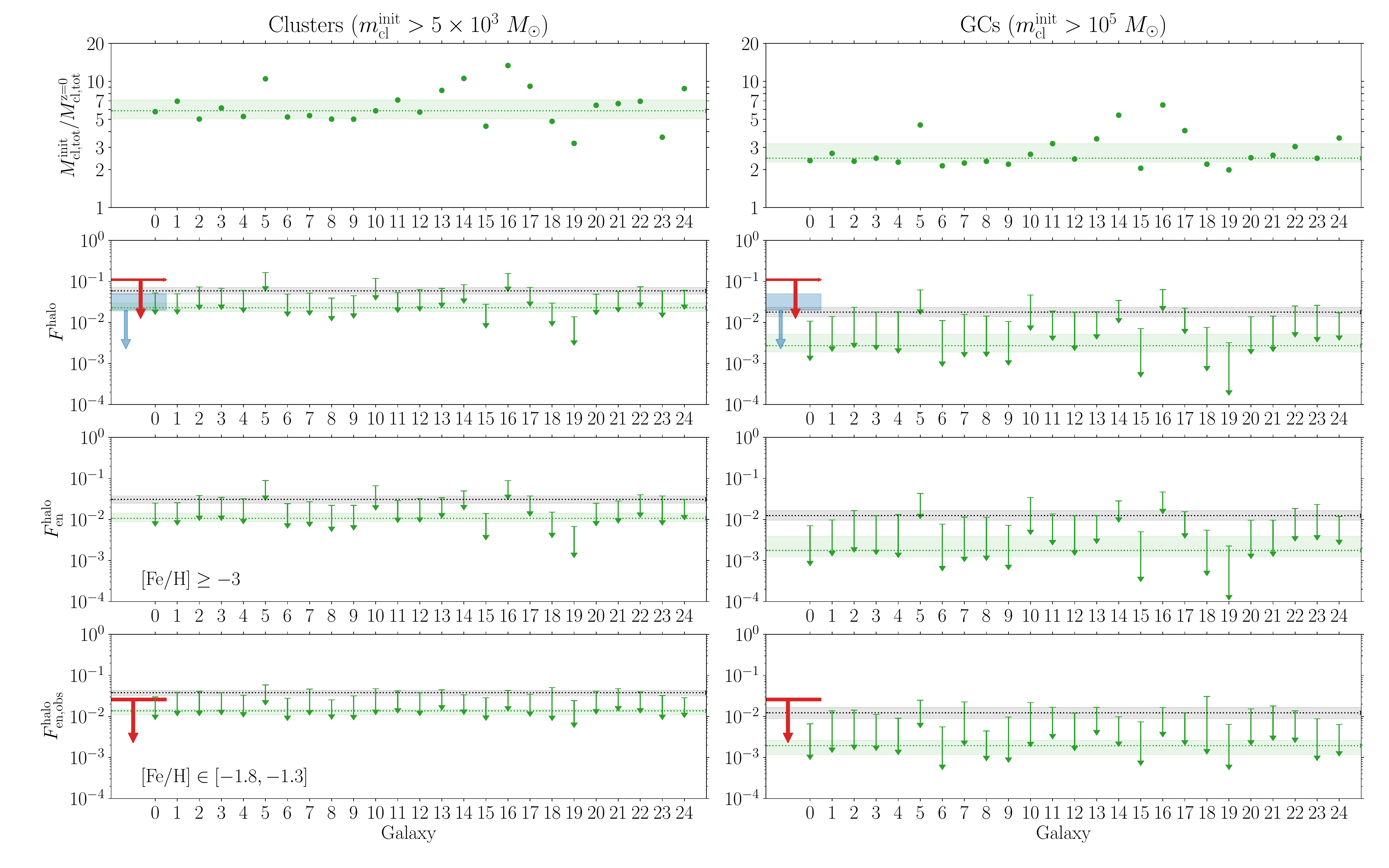}
\caption{\label{fig:gc-contr} Total initial-to-present mass ratios of clusters and GCs (\textit{first row}), mass fraction of halo stars contributed by clusters and GCs (\textit{second row}), mass fraction of halo stars contributed by chemically-distinct stars from clusters and GCs (\textit{third row}), and mass fraction of halo stars contributed by chemically-distinct stars from clusters and GCs matching the metallicity range of \citet{koch19} (\textit{fourth row}) in each of our 25 present-day Milky Way-mass simulations. We define the stellar halo as described in the text, and define the cluster population to be older than $2~\gyr$, more metal-rich than $\rm [Fe/H]> -3$~dex, and part of the stellar halo. The GC population is also restricted to be more massive than $m_{\rm cl}^{\rm init}\geq10^5~\msun$ at birth. In order to mimic observations, we restrict the metallicity range in the fourth row to $\feh\in[-1.8, -1.3]$ for the cluster and GC populations, and for the halo field stars. The upper limits correspond to the extreme case in which no cluster or GC survive to the present day. The horizontal dotted lines and the shaded regions indicate the median and the $25$th--$75$th percentiles, respectively, of the arrow bases (top) and arrow heads (bottom), for each population over our galaxy sample. The red line and downward arrow shown in the second and fourth rows mark the derived upper limit of the mass fraction of halo stars contributed by GCs of $\sim 11$ per cent from \citet{koch19}, and the observationally-inferred fraction of chemically-distinct stars in the Galactic halo \citep[$2.6\pm0.2$ per cent, ][]{koch19}, respectively, whereas the blue box and downward arrow shown in the second row correspond to the revised observational upper limit of $2$--$5$ per cent calculated in this work (see Sect.~\ref{sec:com-frac-mw}).}
\end{figure*}

We determine the total final and initial masses in the cluster and GC populations for each of our simulations, and we show the resulting total initial-to-present mass ratios in the top row of Fig.~\ref{fig:gc-contr}. We find that the total numerically-resolved cluster populations are a median $\sim 5.8$ times more massive at birth, but when restricting to massive clusters, the total initial GC populations are only a median $\sim 2.5$ times more massive than at the present day, in agreement with our earlier findings \citep{reina-campos18}.  

We then calculate the mass fraction of halo field stars contributed by clusters and by GCs as the relative contribution of the dynamically disrupted mass to the mass of the stellar halo,
\be
F^{\rm halo} = \dfrac{\Sigma_{i}^{N_*} \left(M_{{\rm cl}, i}^{\rm init}f_{*, i} - M_{{\rm cl}, i}^{\rm z=0}\right)}{M^{\rm halo}},
\ee
where $M_{{\rm cl}, i}^{\rm init}$ and $M_{{\rm cl}, i}^{\rm z=0}$ correspond to the total initial and final masses of the cluster population contained in the stellar particle $i$, $f_{*} = M_*/M_*^{\rm init}$ is a factor to correct for stellar evolutionary mass loss, and $M^{\rm halo}$ corresponds to the total mass of the halo field stars. Due to the lack of an explicit model for the cold, dense gas of the interstellar medium in \eagle, which is predicted to dominate the disruptive power of galaxies, cluster disruption is underestimated in \emosaics \citep{pfeffer18}. This underdisruption occurs at all gas densities, but it is particularly important at high metallicities ($\feh>-1$~dex) since those clusters do not migrate from their birth environment (see appendix D in \citealt{kruijssen19a}). This implies that the mass fractions calculated using the dynamically disrupted mass from clusters and GCs are a lower limit, and we consider as an upper limit the extreme case in which no cluster or GC survives to the present day, i.e.~all the initial mass in clusters and GCs is disrupted and contributes to the build-up of the halo. We show these fractions for each of our simulations in the second row of Fig.~\ref{fig:gc-contr}. 

We find that clusters contribute a median $2.3$ per cent of mass to the stellar halo, whereas the GC populations present in our suite of simulations contribute a median $0.3$ per cent of the mass in the stellar halo (with the $25$--$75$th percentiles spanning $1.9$--$3.0$ and $0.2$--$0.5$ per cent, respectively). The mass fractions of halo stars contributed from clusters or GCs increase to a median $5.9$ and a median $1.8$ per cent among our suite of galaxies under the extreme assumption that no clusters survive to the present day, respectively, with the $25$--$75$th percentiles spanning $4.9$--$7.6$ and $1.4$--$2.4$ per cent.

The scatter in the halo mass fraction contributed by GCs among our sample is caused by the differences in the formation and assembly history of each host galaxy. Two particularly noteworthy examples are galaxies MW16 and MW19, as they represent two very distinct cases. MW16 undergoes a rich history of mergers, as it is assembled from 38 distinct resolved progenitors with stellar masses $\gtrsim 4 \times 10^6~\msun$ (see table A.3 in \citealt{kruijssen19a}), and exhibits a peak in its GC formation rate and a steep GC age-metallicity relation, which lead to a GC population that is more massive than the median among our galaxy sample, both initially and at the present day. Its rich merger history also leads to a high degree of dynamical disruption. Although the stellar halo of this galaxy is the most massive among our galaxy sample, $M^{\rm halo}\sim10^{10}~\msun$, the mass fraction of halo stars that formed in GCs in this galaxy is the highest among our suite of simulations, $F^{\rm halo} \simeq 2$--$7$ per cent for our two bracketing cases. By contrast, galaxy MW19 forms its stars primarily in-situ and exhibits a shallower age-metallicity relation, and no significant peak in its GC formation rate \citep{kruijssen19a}. This leads to a GC population that is significantly less massive than the median, and a smaller mass fraction of halo stars that formed in GCs, about $0.02$ per cent, which increases to $0.3$ per cent under the extreme assumption that no GCs survive to the present day in the halo. 

\begin{figure*}
\centering
\includegraphics[width=\hsize,keepaspectratio]{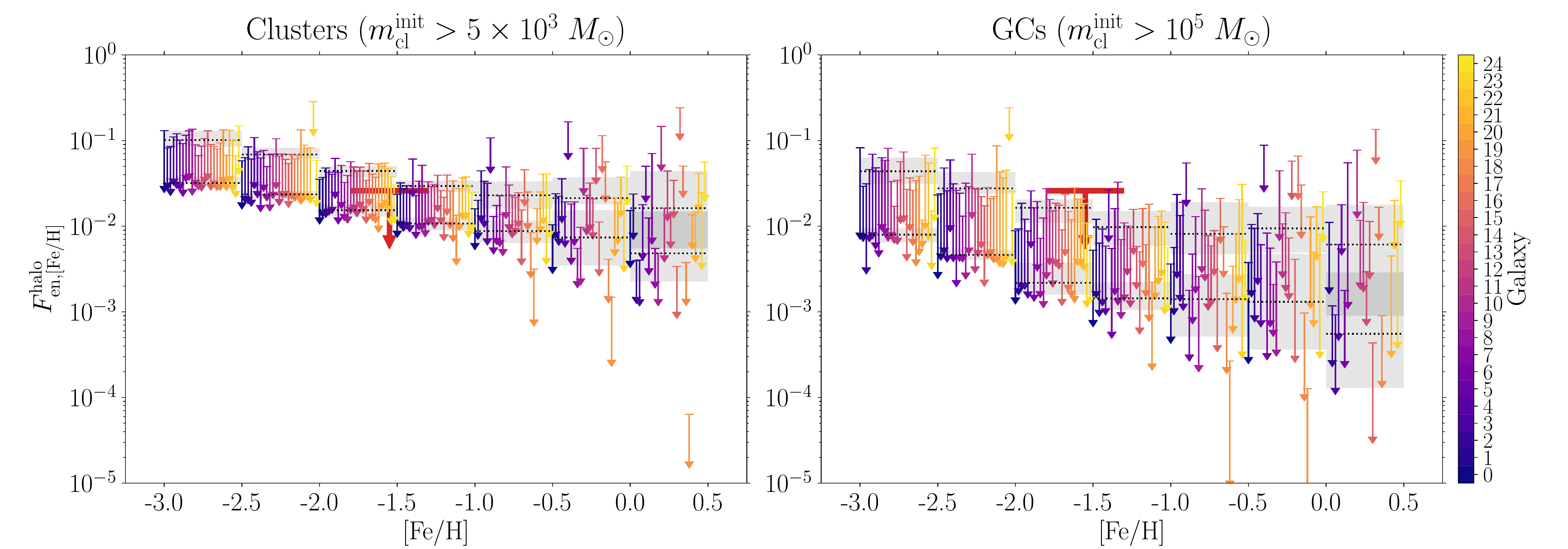}
\caption{\label{fig:enrichedfrac-feh} Mass fractions of halo stars contributed by chemically-distinct stars from clusters (\textit{left panel}) and GCs (\textit{right panel}), as a function of metallicity in each of the 25 present-day Milky Way-mass galaxies of the \emosaics simulations. In order to mimic observational studies, we consider the same metallicity range for both the cluster and GC populations and for the halo field stars. The upper limits correspond to the extreme case in which no cluster or GC survives in the stellar halo. For each metallicity bin, the dotted black lines and grey shaded regions indicate the median and $25$--$75$th percentiles, respectively of the arrow bases (top) and arrow heads (bottom). The red lines and downward arrows correspond to the observationally-inferred fraction of chemically-distinct stars in the Galactic halo \citep[$2.6\pm0.2$ per cent, ][]{koch19}.}
\end{figure*}

We can now use the observed relation between the fraction of chemically-enriched stars and cluster mass from \citet{milone17} to predict the mass fraction of enriched stars contributed by clusters and GCs to the stellar haloes in our suite of simulations. \citet{reina-campos18} suggest that, given the dynamical disruption mechanisms considered in \emosaics, which are postulated to dominate cluster evolution, the observed positive correlation between the enriched fraction and cluster mass likely signifies the initial relation at the time the cluster was born. We thus assume that the observed positive trend describes the initial enriched fraction in our clusters,
\be
f_{\rm en} = 0.189 \log_{10}(m/M_{\odot}) - 0.367,
\ee
with $m = m_{\rm cl}^{\rm init}$ being the initial cluster mass and calculate the total fraction of chemically-distinct mass contributed by our cluster and GC populations to the stellar halo, $F_{\rm en}^{\rm halo}$, which we show in the third row of Fig.~\ref{fig:gc-contr}. We assume there is no preferential mass loss of the unenriched stellar population relative to the chemically-enriched stars within the cluster, so both populations are lost at the same rate. We find that the cluster and GC populations in our suite of galaxies contribute a median $1.1$ and $0.2$ per cent of chemically-distinct mass to the stellar halo, respectively, with the $25$--$75$th percentiles ranging between $0.9$--$1.4$ and $0.1$--$0.4$ per cent. 

We now restrict our cluster and GC populations, as well as our stellar haloes, to the same metallicity range typically used in chemical tagging studies ($\feh\in[-1.8, -1.3]$, e.g.~\citealt{koch19}) and recalculate the total mass fraction of chemically-distinct stars contributed by clusters and GCs to the stellar halo, $F_{\rm en, obs}^{\rm halo}$, which we show in the bottom row of Fig.~\ref{fig:gc-contr}. We find that the medians of the recalculated fraction of chemically-distinct stars do not change significantly relative to the metallicity-unrestricted case ($\feh>-3$~dex, third row in Fig.~\ref{fig:gc-contr}), but the scatter among our galaxy sample decreases.

In order to investigate the influence of the metallicity scale used in \eagle in setting this result, as well as the decrease in the scatter, we explore the dependence of the mass fractions of chemically-distinct halo stars contributed by clusters on the metallicity range considered in Fig.~\ref{fig:enrichedfrac-feh}. We find that, as the metallicity considered increases, the median mass fractions of chemically-distinct halo stars among our suite of galaxies decrease, whereas the scatter in each metallicity bin increases. We also find that an offset of $\pm0.3$ dex in the metallicity scale of the \eagle model would not change the metallicity-limited fractions (fourth row of Fig.~\ref{fig:gc-contr}); at metallicities of $\feh\leq-1$~dex, the mass fractions of chemically-distinct stars in the halo exhibit a normalized interquartile range of $\sim0.4$, which increase to an order of magnitude larger at higher metallicities. Lastly, we find that the increase of the scatter towards large metallicities is caused by the age limit used to define our cluster and GC populations (ages older than $2~\gyr$), which neglects the most recent star formation in the galaxy. As a result, if the observed fraction of enriched stars in the halo is made at the low end of the GC metallicity range, this can lead to an overestimation of the total mass contributed by GCs to the stellar halo.

\section{Comparison to the Milky Way}\label{sec:com-frac-mw}

Observational studies work under the assumption that GCs are a unique site for the formation of chemically distinct stellar populations, and use the chemical signature observed (characterised by a depletion in C, O, and Mg as well as an enhancement in N, Na, and Al, e.g.~\citealt{bastian17}) to estimate the mass fraction contributed by GCs (either currently still bound or fully disrupted) to the stellar halo of the Milky Way \citep[e.g.][]{martell10,martell11,martell16,koch19}. A number of surveys have looked for stars with anomalous chemistry in samples of halo stars, mainly through either N or Na enhancement. Such an estimate naturally only corresponds to the contribution to the halo from clusters that host multiple populations, the presence of which appears to be related to the initial mass of the cluster, near $\sim10^5 M_{\odot}$ \citep{kruijssen15b,reina-campos18}. The results of these different approaches have been quite consistent, with authors finding between $1.4$--$2.6$ per cent of halo stars showing N or Na-enhancement \citep[e.g.][]{martell10,carretta10b,koch19}.

Given that low-mass stars dominate the Chabrier stellar initial mass function used in \emosaics by number, we convert the obtained mass fractions to number fractions by assuming a one-to-one conversion and compare our results with those obtained using the chemical-tagging technique. This way, we find that the metallicity-limited fractions of chemically-distinct stars in the halo contributed by disrupted GCs (bottom right panel in Fig.~\ref{fig:gc-contr}) estimated in this work are consistent with the observational estimates \citep[e.g.][]{martell16,koch19}, although perhaps a bit low. In contrast, assuming that chemically-distinct stars can also form in low-mass stellar clusters (more massive than $5\times10^3~\msun$ at birth), we find that the mass fraction of chemically-distinct stars in the halo exhibit better agreement with the observational results (bottom left panel in Fig.~\ref{fig:gc-contr}). This suggests that fully-disrupted low-mass stellar clusters might also exhibit stars with light-element inhomogenities that are contributed to the stellar halo when the cluster dissolves.

In order to find the total contribution of GCs to the stellar halo, this fraction needs to be corrected for the unseen primordial or unenriched stars with field-like abundances that are not detectable in such chemical-tagging surveys.  In earlier works, this was done by adopting the heavy mass-loss invoked in multiple population formation models \citep[e.g.][]{dercole08,schaerer11} in order to solve the ``mass budget problem" (see \citealt{bastian17} for a recent review). This correction factor was largely unconstrained, resulting in estimates between $17$--$40$ per cent for the mass fraction contribution of GC stars to the halo.

\citet{koch19} adopt a more physically motivated formalism to estimate the correction factor, using the constraint that GCs were (on average) only a factor two more massive than at present, as derived through comparisons of the index of the low-mass stellar mass function within GCs with cluster disruption models \citep[e.g.][]{kruijssen09,webb15}. The authors also assume that all of the chemically-enriched stars present in the halo are contributed from fully disrupted GCs, so that the existing GC population only lost unenriched stars to the halo. In addition, the authors assume that all GCs, regardless of their metallicity and orbit, contribute to the stellar halo. Under these assumptions, the authors estimate an upper limit of $11$ per cent of the stellar halo is made up of stars formed originally in GCs.

However, \citet{forbes18} have estimated the total mass in existing GCs in the stellar halo, $2.6 \times 10^7~M_{\odot}$ \citep[see also][]{kruijssen09b}, as well as the total mass lost by each of these GCs ($2.5\times10^5 M_{\odot}$ on average, and a total of $2.5\times10^7 M_{\odot}$ for the $\sim100$ halo GCs considered). They compared that to the total mass of the Galactic stellar halo ($1.5\pm0.4\times10^9~\msun$; \citealt{deason19b}) and found that the total mass of enriched stars presently in GCs is $\sim1.3 \times10^7 M_{\odot}$ (assuming an enriched-to-total fraction of $50$ per cent; this increases slightly if more realistic values, $f_{\rm en}=0.67$, are used). The fact that this is very close to the observed mass in halo field enriched stars ($1.4$--$2.6$ per cent $\times (1.5\pm0.4)\times10^9~\msun$ = $1.5$--$4.9\times10^7 M_{\odot})$ suggests that the amount of mass loss from existing GCs is enough to explain the observed number of enriched stars in the halo under the assumption that both unenriched and enriched stars are lost at similar rates \citep[also see][]{kruijssen15b}. This suggests that the original number of GCs was of the similar order of magnitude as the current one.

Such a conclusion is also supported by studies that have found that the present day fraction of enriched stars in GCs is representative of the initial fraction \citep[e.g][]{larsen12,bastian15,reina-campos18}.  If each present-day GC had a factor of $2$--$4$~times more unenriched stars at birth than they currently do, this would represent a drastic difference from their present day values and would be inconsistent with a number of observations and expectations \citep[e.g.][]{bastian17}.

Under the assumption that the current population ratio, i.e.~enriched-to-unenriched, is similar to the initial one, we would only need to correct the observed fraction of chemically-distinct stars in the halo ($1.4$--$2.6$ per cent) for the population ratio between unenriched and enriched stars.  Adopting a 50/50 ratio leads to $2.8$--$5.2$ per cent, while adopting an enriched fraction of $f_{\rm en} = 0.67$ \citep{milone17} leads to fractions of $2.1$--$3.9$ per cent of the Galactic stellar halo being contributed by GCs\footnote{Defined as stellar clusters that host multiple populations, which we assume are those older than $2~\gyr$, more metal-rich than $\feh>-3$ dex and more massive than $m_{\rm cl}^{\rm init} \geq 10^5~\msun$ at birth.}. This estimate remains unchanged if one posits that disrupted GCs (as opposed to dissolving GCs that still exist) are the main contributor of GC stars to the halo, as long as the enriched-to-unenriched ratio was similar in these clusters to that of existing GCs.

Finally, we note, following \citet{koch19}, that these estimates are upper limits, as other processes (e.g. binary evolution) can lead to normal stars appearing as enriched stars. Hence, the likely contribution of GCs to the stellar halo is lower than the $2$--$5$ per cent estimated here.

We can now compare these observational estimates to the mass fractions of halo stars dynamically lost from clusters and GCs in the \emosaics simulations, which are shown in the second row of Fig.~\ref{fig:gc-contr}. We find that the simulated fractions of halo stars contributed by GCs are consistent with the revised observational upper limits of $2$--$5$ per cent of the Galactic halo stars originating in GCs, implying that GCs play a sub-dominant role in the build-up of stellar haloes.

This result is in agreement with recent observational studies that suggest that the bulk of the Galactic stellar halo is assembled from tidally-disrupted dwarf galaxies. \citet{deason15} argue that the relatively high ratio of blue stragglers to blue horizontal branch stars in the stellar halo is inconsistent with the low ratios observed in GCs, and suggests a scenario in which massive dwarfs are the dominant building blocks of the Galactic stellar halo. Moreover, a large number of studies using data from the \textit{Gaia} mission suggest that the accretion of a single massive ($M\sim10^9~\msun$) satellite $\sim 9$--$10~\gyr$ ago could be the origin of the inner Galactic stellar halo (\textit{Gaia}-Enceladus/Sausage; e.g. \citealt{belokurov18,helmi18}; also see \citealt{kruijssen19b}). In addition, recent observations of the Galactic halo find that the majority of the halo is composed by discrete features in orbital-chemical space, indicating that the bulk of the halo (or specifically $\sim 70$~per cent, see \citealt{mackereth19}) has assembled from the accretion of tidally-disrupted dwarfs galaxies \citep{conroy19}. Thus, the small fractions of halo stars contributed by GCs calculated in this work using the \emosaics simulations would also favour this formation scenario of the Galactic stellar halo.

\section{Summary}

We use the 25 present-day Milky Way-mass cosmological zoom simulations from the \emosaics project to quantify the total mass fraction, as well as the chemically-distinct mass fraction, contributed to the stellar halo by clusters and GCs, and compare the results with recent observations.

We define our cluster and GC populations to resemble those in which light-element abundance variations have been observed. Thus, the halo cluster populations are defined to be older than $2~\gyr$ and more metal-rich than $\feh>-3~{\rm dex}$. In order to reduce memory requirements in \emosaics, only clusters initially more massive than $m_{\rm cl}^{\rm init}\geq5\times10^3~\msun$ are evolved and considered in the numerically-resolved cluster populations. In addition to these criteria, we consider as halo GCs those clusters more massive than $m_{\rm cl}^{\rm init}\geq10^5~\msun$ at birth.

We find that the stellar haloes in our central galaxies contain a median $2.3$ and $0.3$ per cent of mass that formed as part of a cluster or a GC, with the $25$--$75$th percentiles spanning $1.9$--$3.0$ and $0.2$--$0.5$ per cent, respectively. The scatter among our galaxy sample can be traced to differences in the assembly histories of the host galaxies (see Sect.~\ref{sec:frac-emosaics}). Using the observed positive correlation between the fraction of enriched stars and their cluster mass from \citet{milone17}, we determine the mass fraction of the stellar halo contributed from disrupted clusters and GCs that would exhibit light-element abundance variations. We find that among our suite of galaxies, there is a median $1.1$ and $0.2$ per cent of mass in the stellar halo that is chemically enriched contributed by clusters and GCs, respectively. These small fractions imply that clusters and GCs play a sub-dominant role in the build-up of the stellar halo. This result is in agreement with recent studies that suggest that the Galactic stellar halo has assembled from tidally-disrupted dwarf galaxies \citep[e.g.][]{deason15,belokurov18,helmi18,conroy19,kruijssen19b}.

We also find that the mass fraction of chemically-enriched stars in the halo contributed by clusters and GCs depends on the metallicity range considered, with decreasing fractions towards higher metallicity bins. As a result, if the observed fraction of enriched stars in the halo is made at the low end of the GC metallicity range, this can lead to an overestimation of the total mass contributed by GCs to the stellar halo.

Comparing our results to recent observational surveys, which use a chemical-tagging technique to identify chemically-distinct stars in the halo and find typical upper limits between $1.4$--$2.6$ per cent \citep[e.g.][]{martell16,koch19}, we find that our predicted fractions of chemically-distinct stars in the halo contributed by GCs are consistent with observations, although perhaps a bit low (fourth row in Fig.~\ref{fig:gc-contr}). This suggests that the amount of mass loss from surviving GCs is enough to explain the observed number of enriched stars in the halo under the assumption that both unenriched and enriched stars are lost at similar rates. This result, in conjunction with the agreement between the properties of observed cluster populations and those simulated in the \emosaics suite of simulations \citep{pfeffer18,kruijssen19a,usher18,pfeffer19b}, suggests that models other than those requiring strong mass loss of unenriched stars are required to describe the formation of chemically enriched stellar populations in GCs. 

\section*{Acknowledgements}
MRC is supported by a Fellowship from the International Max Planck Research School for Astronomy and Cosmic Physics at the University of Heidelberg (IMPRS-HD). MRC and JMDK gratefully acknowledge funding from the European Research Council (ERC) under the European Union's Horizon 2020 research and innovation programme via the ERC Starting Grant MUSTANG (grant agreement number 714907). JMDK gratefully acknowledges funding from the German Research Foundation (DFG) in the form of an Emmy Noether Research Group (grant number KR4801/1-1). JP and NB gratefully acknowledge funding from the ERC under the European Union's Horizon 2020 research and innovation programme via the ERC Consolidator Grant Multi-Pop (grant agreement number 646928). NB and RAC are Royal Society University Research Fellows. EKG and AK acknowledge financial support from the Sonderforschungsbereich SFB 881 ``The Milky Way System'' (subprojects A03, A05, A08) of the DFG. This work used the DiRAC Data Centric system at Durham University, operated by the Institute for Computational Cosmology on behalf of the STFC DiRAC HPC Facility (www.dirac.ac.uk). This equipment was funded by BIS National E-infrastructure capital grant ST/K00042X/1, STFC capital grants ST/H008519/1 and ST/K00087X/1, STFC DiRAC Operations grant ST/K003267/1 and Durham University. DiRAC is part of the National E-Infrastructure. The work also made use of high performance computing facilities at Liverpool John Moores University, partly funded by the Royal Society and LJMU’s Faculty of Engineering and Technology.

\bibliographystyle{mnras}
\bibliography{./bibdesk-bib}

\begin{thebibliography}{}
\makeatletter
\relax
\def\mn@urlcharsother{\let\do\@makeother \do\$\do\&\do\#\do\^\do\_\do\%\do\~}
\def\mn@doi{\begingroup\mn@urlcharsother \@ifnextchar [ {\mn@doi@}
  {\mn@doi@[]}}
\def\mn@doi@[#1]#2{\def\@tempa{#1}\ifx\@tempa\@empty \href
  {http://dx.doi.org/#2} {doi:#2}\else \href {http://dx.doi.org/#2} {#1}\fi
  \endgroup}
\def\mn@eprint#1#2{\mn@eprint@#1:#2::\@nil}
\def\mn@eprint@arXiv#1{\href {http://arxiv.org/abs/#1} {{\tt arXiv:#1}}}
\def\mn@eprint@dblp#1{\href {http://dblp.uni-trier.de/rec/bibtex/#1.xml}
  {dblp:#1}}
\def\mn@eprint@#1:#2:#3:#4\@nil{\def\@tempa {#1}\def\@tempb {#2}\def\@tempc
  {#3}\ifx \@tempc \@empty \let \@tempc \@tempb \let \@tempb \@tempa \fi \ifx
  \@tempb \@empty \def\@tempb {arXiv}\fi \@ifundefined
  {mn@eprint@\@tempb}{\@tempb:\@tempc}{\expandafter \expandafter \csname
  mn@eprint@\@tempb\endcsname \expandafter{\@tempc}}}

\bibitem[\protect\citeauthoryear{{Adamo}, {Kruijssen}, {Bastian}, {Silva-Villa}
   \& {Ryon}}{{Adamo} et~al.}{2015}]{adamo15b}
{Adamo} A.,  {Kruijssen} J.~M.~D.,  {Bastian} N.,  {Silva-Villa} E.,   {Ryon}
  J.,  2015, \mn@doi [\mnras] {10.1093/mnras/stv1203}, \href
  {http://adsabs.harvard.edu/abs/2015MNRAS.452..246A} {452, 246}

\bibitem[\protect\citeauthoryear{{Bastian}}{{Bastian}}{2008}]{bastian08}
{Bastian} N.,  2008, \mn@doi [\mnras] {10.1111/j.1365-2966.2008.13775.x}, \href
  {http://adsabs.harvard.edu/abs/2008MNRAS.390..759B} {390, 759}

\bibitem[\protect\citeauthoryear{{Bastian} \& {Lardo}}{{Bastian} \&
  {Lardo}}{2015}]{bastian15}
{Bastian} N.,  {Lardo} C.,  2015, \mn@doi [\mnras] {10.1093/mnras/stv1661},
  \href {http://adsabs.harvard.edu/abs/2015MNRAS.453..357B} {453, 357}

\bibitem[\protect\citeauthoryear{Bastian \& Lardo}{Bastian \&
  Lardo}{2018}]{bastian17}
Bastian N.,  Lardo C.,  2018, \mn@doi [\araa]
  {10.1146/annurev-astro-081817-051839}, 56

\bibitem[\protect\citeauthoryear{{Belokurov}, {Erkal}, {Evans}, {Koposov}  \&
  {Deason}}{{Belokurov} et~al.}{2018}]{belokurov18}
{Belokurov} V.,  {Erkal} D.,  {Evans} N.~W.,  {Koposov} S.~E.,   {Deason}
  A.~J.,  2018, \mn@doi [\mnras] {10.1093/mnras/sty982}, \href
  {https://ui.adsabs.harvard.edu/abs/2018MNRAS.478..611B} {478, 611}

\bibitem[\protect\citeauthoryear{{Carollo}, {Martell}, {Beers}  \&
  {Freeman}}{{Carollo} et~al.}{2013}]{carollo13}
{Carollo} D.,  {Martell} S.~L.,  {Beers} T.~C.,   {Freeman} K.~C.,  2013,
  \mn@doi [\apj] {10.1088/0004-637X/769/2/87}, \href
  {https://ui.adsabs.harvard.edu/\#abs/2013ApJ...769...87C} {769, 87}

\bibitem[\protect\citeauthoryear{{Carretta}}{{Carretta}}{2016}]{carretta16}
{Carretta} E.,  2016, in {Bragaglia} A.,  {Arnaboldi} M.,  {Rejkuba} M.,
  {Romano} D.,  eds,  IAU Symposium Vol. 317, The General Assembly of Galaxy
  Halos: Structure, Origin and Evolution. pp 97--103 (\mn@eprint {arXiv}
  {1510.00507}), \mn@doi{10.1017/S1743921315006730}

\bibitem[\protect\citeauthoryear{{Carretta}, {Bragaglia}, {Gratton}  \&
  {Lucatello}}{{Carretta} et~al.}{2009}]{carretta09a}
{Carretta} E.,  {Bragaglia} A.,  {Gratton} R.,   {Lucatello} S.,  2009, \mn@doi
  [\aap] {10.1051/0004-6361/200912097}, \href
  {http://adsabs.harvard.edu/abs/2009A%26A...505..139C} {505, 139}

\bibitem[\protect\citeauthoryear{{Carretta}, {Bragaglia}, {Gratton},
  {Recio-Blanco}, {Lucatello}, {D'Orazi}  \& {Cassisi}}{{Carretta}
  et~al.}{2010}]{carretta10b}
{Carretta} E.,  {Bragaglia} A.,  {Gratton} R.~G.,  {Recio-Blanco} A.,
  {Lucatello} S.,  {D'Orazi} V.,   {Cassisi} S.,  2010, \mn@doi [\aap]
  {10.1051/0004-6361/200913451}, \href
  {http://adsabs.harvard.edu/abs/2010A%26A...516A..55C} {516, A55}

\bibitem[\protect\citeauthoryear{{Conroy}, {Naidu}, {Zaritsky}, {Bonaca},
  {Cargile}, {Johnson}  \& {Caldwell}}{{Conroy} et~al.}{2019}]{conroy19}
{Conroy} C.,  {Naidu} R.~P.,  {Zaritsky} D.,  {Bonaca} A.,  {Cargile} P.,
  {Johnson} B.~D.,   {Caldwell} N.,  2019, arXiv e-prints, \href
  {https://ui.adsabs.harvard.edu/abs/2019arXiv190902007C} {p. arXiv:1909.02007}

\bibitem[\protect\citeauthoryear{{Crain} et~al.,}{{Crain}
  et~al.}{2015}]{crain15}
{Crain} R.~A.,  et~al., 2015, \mn@doi [\mnras] {10.1093/mnras/stv725}, \href
  {http://adsabs.harvard.edu/abs/2015MNRAS.450.1937C} {450, 1937}

\bibitem[\protect\citeauthoryear{{D'Ercole}, {Vesperini}, {D'Antona},
  {McMillan}  \& {Recchi}}{{D'Ercole} et~al.}{2008}]{dercole08}
{D'Ercole} A.,  {Vesperini} E.,  {D'Antona} F.,  {McMillan} S.~L.~W.,
  {Recchi} S.,  2008, \mn@doi [\mnras] {10.1111/j.1365-2966.2008.13915.x},
  \href {http://adsabs.harvard.edu/abs/2008MNRAS.391..825D} {391, 825}

\bibitem[\protect\citeauthoryear{{Deason}, {Belokurov}  \& {Weisz}}{{Deason}
  et~al.}{2015}]{deason15}
{Deason} A.~J.,  {Belokurov} V.,   {Weisz} D.~R.,  2015, \mn@doi [\mnras]
  {10.1093/mnrasl/slv001}, \href
  {https://ui.adsabs.harvard.edu/abs/2015MNRAS.448L..77D} {448, L77}

\bibitem[\protect\citeauthoryear{{Deason}, {Belokurov}  \& {Sanders}}{{Deason}
  et~al.}{2019}]{deason19b}
{Deason} A.~J.,  {Belokurov} V.,   {Sanders} J.~L.,  2019, arXiv e-prints,
  \href {https://ui.adsabs.harvard.edu/abs/2019arXiv190802763D} {p.
  arXiv:1908.02763}

\bibitem[\protect\citeauthoryear{{Forbes} et~al.,}{{Forbes}
  et~al.}{2018}]{forbes18}
{Forbes} D.~A.,  et~al., 2018, \mn@doi [Proceedings of the Royal Society of
  London Series A] {10.1098/rspa.2017.0616}, \href
  {http://adsabs.harvard.edu/abs/2018RSPSA.47470616F} {474, 20170616}

\bibitem[\protect\citeauthoryear{{Helmi}, {Babusiaux}, {Koppelman}, {Massari},
  {Veljanoski}  \& {Brown}}{{Helmi} et~al.}{2018}]{helmi18}
{Helmi} A.,  {Babusiaux} C.,  {Koppelman} H.~H.,  {Massari} D.,  {Veljanoski}
  J.,   {Brown} A. G.~A.,  2018, \mn@doi [\nat] {10.1038/s41586-018-0625-x},
  \href {https://ui.adsabs.harvard.edu/abs/2018Natur.563...85H} {563, 85}

\bibitem[\protect\citeauthoryear{{Hughes}, {Pfeffer}, {Martig}, {Reina-Campos},
  {Bastian}, {Crain}  \& {Kruijssen}}{{Hughes} et~al.}{2019a}]{hughes19b}
{Hughes} M.~E.,  {Pfeffer} J.,  {Martig} M.,  {Reina-Campos} M.,  {Bastian} N.,
   {Crain} R.~A.,   {Kruijssen} J.~M.~D.,  2019a, \mnras~submitted

\bibitem[\protect\citeauthoryear{{Hughes}, {Pfeffer}, {Martig}, {Bastian},
  {Crain}, {Kruijssen}  \& {Reina-Campos}}{{Hughes} et~al.}{2019b}]{hughes19}
{Hughes} M.~E.,  {Pfeffer} J.,  {Martig} M.,  {Bastian} N.,  {Crain} R.~A.,
  {Kruijssen} J.~M.~D.,   {Reina-Campos} M.,  2019b, \mn@doi [\mnras]
  {10.1093/mnras/sty2889}, \href
  {https://ui.adsabs.harvard.edu/\#abs/2019MNRAS.482.2795H} {482, 2795}

\bibitem[\protect\citeauthoryear{{Johnson} et~al.,}{{Johnson}
  et~al.}{2016}]{johnson16}
{Johnson} L.~C.,  et~al., 2016, \mn@doi [\apj] {10.3847/0004-637X/827/1/33},
  \href {http://adsabs.harvard.edu/abs/2016ApJ...827...33J} {827, 33}

\bibitem[\protect\citeauthoryear{{Koch}, {Grebel}  \& {Martell}}{{Koch}
  et~al.}{2019}]{koch19}
{Koch} A.,  {Grebel} E.~K.,   {Martell} S.~L.,  2019, \mn@doi [\aap]
  {10.1051/0004-6361/201834825}, \href
  {https://ui.adsabs.harvard.edu/abs/2019A&A...625A..75K} {625, A75}

\bibitem[\protect\citeauthoryear{{Kruijssen}}{{Kruijssen}}{2012}]{kruijssen12d}
{Kruijssen} J.~M.~D.,  2012, \mn@doi [\mnras]
  {10.1111/j.1365-2966.2012.21923.x}, \href
  {http://adsabs.harvard.edu/abs/2012MNRAS.426.3008K} {426, 3008}

\bibitem[\protect\citeauthoryear{{Kruijssen}}{{Kruijssen}}{2015}]{kruijssen15b}
{Kruijssen} J.~M.~D.,  2015, \mn@doi [\mnras] {10.1093/mnras/stv2026}, \href
  {http://adsabs.harvard.edu/abs/2015MNRAS.454.1658K} {454, 1658}

\bibitem[\protect\citeauthoryear{{Kruijssen} \& {Mieske}}{{Kruijssen} \&
  {Mieske}}{2009}]{kruijssen09}
{Kruijssen} J.~M.~D.,  {Mieske} S.,  2009, \mn@doi [\aap]
  {10.1051/0004-6361/200811453}, \href
  {http://adsabs.harvard.edu/abs/2009A%26A...500..785K} {500, 785}

\bibitem[\protect\citeauthoryear{{Kruijssen} \& {Portegies Zwart}}{{Kruijssen}
  \& {Portegies Zwart}}{2009}]{kruijssen09b}
{Kruijssen} J.~M.~D.,  {Portegies Zwart} S.~F.,  2009, \mn@doi [\apjl]
  {10.1088/0004-637X/698/2/L158}, \href
  {https://ui.adsabs.harvard.edu/abs/2009ApJ...698L.158K} {698, L158}

\bibitem[\protect\citeauthoryear{{Kruijssen}, {Pelupessy}, {Lamers}, {Portegies
  Zwart}  \& {Icke}}{{Kruijssen} et~al.}{2011}]{kruijssen11}
{Kruijssen} J.~M.~D.,  {Pelupessy} F.~I.,  {Lamers} H.~J.~G.~L.~M.,  {Portegies
  Zwart} S.~F.,   {Icke} V.,  2011, \mn@doi [\mnras]
  {10.1111/j.1365-2966.2011.18467.x}, \href
  {http://adsabs.harvard.edu/abs/2011MNRAS.414.1339K} {414, 1339}

\bibitem[\protect\citeauthoryear{{Kruijssen}, {Pfeffer}, {Crain}  \&
  {Bastian}}{{Kruijssen} et~al.}{2019a}]{kruijssen19a}
{Kruijssen} J.~M.~D.,  {Pfeffer} J.~L.,  {Crain} R.~A.,   {Bastian} N.,  2019a,
  \mn@doi [\mnras] {10.1093/mnras/stz968}, \href
  {https://ui.adsabs.harvard.edu/abs/2019MNRAS.486.3134K} {486, 3134}

\bibitem[\protect\citeauthoryear{{Kruijssen}, {Pfeffer}, {Reina-Campos},
  {Crain}  \& {Bastian}}{{Kruijssen} et~al.}{2019b}]{kruijssen19b}
{Kruijssen} J.~M.~D.,  {Pfeffer} J.~L.,  {Reina-Campos} M.,  {Crain} R.~A.,
  {Bastian} N.,  2019b, \mn@doi [\mnras] {10.1093/mnras/sty1609}, \href
  {https://ui.adsabs.harvard.edu/abs/2019MNRAS.486.3180K} {486, 3180}

\bibitem[\protect\citeauthoryear{{Larsen}, {Strader}  \& {Brodie}}{{Larsen}
  et~al.}{2012}]{larsen12}
{Larsen} S.~S.,  {Strader} J.,   {Brodie} J.~P.,  2012, \mn@doi [\aap]
  {10.1051/0004-6361/201219897}, \href
  {http://adsabs.harvard.edu/abs/2012A%26A...544L..14L} {544, L14}

\bibitem[\protect\citeauthoryear{{Mackereth} \& {Bovy}}{{Mackereth} \&
  {Bovy}}{2019}]{mackereth19}
{Mackereth} J.~T.,  {Bovy} J.,  2019, arXiv e-prints, \href
  {https://ui.adsabs.harvard.edu/abs/2019arXiv191003590M} {p. arXiv:1910.03590}

\bibitem[\protect\citeauthoryear{{Martell} \& {Grebel}}{{Martell} \&
  {Grebel}}{2010}]{martell10}
{Martell} S.~L.,  {Grebel} E.~K.,  2010, \mn@doi [\aap]
  {10.1051/0004-6361/201014135}, \href
  {https://ui.adsabs.harvard.edu/\#abs/2010A&A...519A..14M} {519, A14}

\bibitem[\protect\citeauthoryear{{Martell}, {Smolinski}, {Beers}  \&
  {Grebel}}{{Martell} et~al.}{2011}]{martell11}
{Martell} S.~L.,  {Smolinski} J.~P.,  {Beers} T.~C.,   {Grebel} E.~K.,  2011,
  \mn@doi [\aap] {10.1051/0004-6361/201117644}, \href
  {https://ui.adsabs.harvard.edu/\#abs/2011A&A...534A.136M} {534, A136}

\bibitem[\protect\citeauthoryear{{Martell} et~al.,}{{Martell}
  et~al.}{2016}]{martell16}
{Martell} S.~L.,  et~al., 2016, \mn@doi [\apj] {10.3847/0004-637X/825/2/146},
  \href {https://ui.adsabs.harvard.edu/\#abs/2016ApJ...825..146M} {825, 146}

\bibitem[\protect\citeauthoryear{{Martocchia} et~al.,}{{Martocchia}
  et~al.}{2018}]{martocchia18}
{Martocchia} S.,  et~al., 2018, \mn@doi [\mnras] {10.1093/mnras/stx2556}, \href
  {https://ui.adsabs.harvard.edu/\#abs/2018MNRAS.473.2688M} {473, 2688}

\bibitem[\protect\citeauthoryear{{Massari}, {Koppelman}  \& {Helmi}}{{Massari}
  et~al.}{2019}]{massari19}
{Massari} D.,  {Koppelman} H.~H.,   {Helmi} A.,  2019, \mn@doi [\aap]
  {10.1051/0004-6361/201936135}, \href
  {https://ui.adsabs.harvard.edu/abs/2019A&A...630L...4M} {630, L4}

\bibitem[\protect\citeauthoryear{{Messa} et~al.,}{{Messa}
  et~al.}{2018}]{messa18}
{Messa} M.,  et~al., 2018, \mn@doi [\mnras] {10.1093/mnras/sty577}, \href
  {http://adsabs.harvard.edu/abs/2018MNRAS.477.1683M} {477, 1683}

\bibitem[\protect\citeauthoryear{{Milone} et~al.,}{{Milone}
  et~al.}{2017}]{milone17}
{Milone} A.~P.,  et~al., 2017, \mn@doi [\mnras] {10.1093/mnras/stw2531}, \href
  {http://adsabs.harvard.edu/abs/2017MNRAS.464.3636M} {464, 3636}

\bibitem[\protect\citeauthoryear{{Myeong}, {Evans}, {Belokurov}, {Sand ers}  \&
  {Koposov}}{{Myeong} et~al.}{2018}]{myeong18d}
{Myeong} G.~C.,  {Evans} N.~W.,  {Belokurov} V.,  {Sand ers} J.~L.,   {Koposov}
  S.~E.,  2018, \mn@doi [\apjl] {10.3847/2041-8213/aad7f7}, \href
  {https://ui.adsabs.harvard.edu/abs/2018ApJ...863L..28M} {863, L28}

\bibitem[\protect\citeauthoryear{{Pfeffer}, {Kruijssen}, {Crain}  \&
  {Bastian}}{{Pfeffer} et~al.}{2018}]{pfeffer18}
{Pfeffer} J.,  {Kruijssen} J.~M.~D.,  {Crain} R.~A.,   {Bastian} N.,  2018,
  \mn@doi [\mnras] {10.1093/mnras/stx3124}, \href
  {http://adsabs.harvard.edu/abs/2018MNRAS.475.4309P} {475, 4309}

\bibitem[\protect\citeauthoryear{{Pfeffer}, {Bastian}, {Crain}, {Diederik
  Kruijssen}, {Hughes}  \& {Reina-Campos}}{{Pfeffer}
  et~al.}{2019a}]{pfeffer19a}
{Pfeffer} J.,  {Bastian} N.,  {Crain} R.~A.,  {Diederik Kruijssen} J.~M.,
  {Hughes} M.~E.,   {Reina-Campos} M.,  2019a, \mn@doi [\mnras]
  {10.1093/mnras/stz1592}, \href
  {https://ui.adsabs.harvard.edu/abs/2019MNRAS.tmp.1526P} {p.~1526}

\bibitem[\protect\citeauthoryear{{Pfeffer}, {Bastian}, {Kruijssen},
  {Reina-Campos}, {Crain}  \& {Usher}}{{Pfeffer} et~al.}{2019b}]{pfeffer19b}
{Pfeffer} J.,  {Bastian} N.,  {Kruijssen} J.~M.~D.,  {Reina-Campos} M.,
  {Crain} R.~A.,   {Usher} C.,  2019b, \mnras, \href
  {https://ui.adsabs.harvard.edu/abs/2019arXiv190710118P} {p. arXiv:1907.10118}

\bibitem[\protect\citeauthoryear{{Piotto} et~al.,}{{Piotto}
  et~al.}{2015}]{piotto15}
{Piotto} G.,  et~al., 2015, \mn@doi [\aj] {10.1088/0004-6256/149/3/91}, \href
  {http://adsabs.harvard.edu/abs/2015AJ....149...91P} {149, 91}

\bibitem[\protect\citeauthoryear{{Reina-Campos} \& {Kruijssen}}{{Reina-Campos}
  \& {Kruijssen}}{2017}]{reina-campos17}
{Reina-Campos} M.,  {Kruijssen} J.~M.~D.,  2017, \mn@doi [\mnras]
  {10.1093/mnras/stx790}, \href
  {http://adsabs.harvard.edu/abs/2017MNRAS.469.1282R} {469, 1282}

\bibitem[\protect\citeauthoryear{{Reina-Campos}, {Kruijssen}, {Pfeffer},
  {Bastian}  \& {Crain}}{{Reina-Campos} et~al.}{2018}]{reina-campos18}
{Reina-Campos} M.,  {Kruijssen} J.~M.~D.,  {Pfeffer} J.,  {Bastian} N.,
  {Crain} R.~A.,  2018, \mn@doi [\mnras] {10.1093/mnras/sty2451}, \href
  {http://adsabs.harvard.edu/abs/2018MNRAS.481.2851R} {481, 2851}

\bibitem[\protect\citeauthoryear{{Reina-Campos}, {Kruijssen}, {Pfeffer},
  {Bastian}  \& {Crain}}{{Reina-Campos} et~al.}{2019}]{reina-campos19}
{Reina-Campos} M.,  {Kruijssen} J.~M.~D.,  {Pfeffer} J.~L.,  {Bastian} N.,
  {Crain} R.~A.,  2019, \mn@doi [\mnras] {10.1093/mnras/stz1236}, \href
  {https://ui.adsabs.harvard.edu/abs/2019MNRAS.486.5838R} {486, 5838}

\bibitem[\protect\citeauthoryear{{Sales}, {Navarro}, {Theuns}, {Schaye},
  {White}, {Frenk}, {Crain}  \& {Dalla Vecchia}}{{Sales}
  et~al.}{2012}]{sales12}
{Sales} L.~V.,  {Navarro} J.~F.,  {Theuns} T.,  {Schaye} J.,  {White} S. D.~M.,
   {Frenk} C.~S.,  {Crain} R.~A.,   {Dalla Vecchia} C.,  2012, \mn@doi [\mnras]
  {10.1111/j.1365-2966.2012.20975.x}, \href
  {https://ui.adsabs.harvard.edu/\#abs/2012MNRAS.423.1544S} {423, 1544}

\bibitem[\protect\citeauthoryear{{Schaerer} \& {Charbonnel}}{{Schaerer} \&
  {Charbonnel}}{2011}]{schaerer11}
{Schaerer} D.,  {Charbonnel} C.,  2011, \mn@doi [\mnras]
  {10.1111/j.1365-2966.2011.18304.x}, \href
  {http://adsabs.harvard.edu/abs/2011MNRAS.413.2297S} {413, 2297}

\bibitem[\protect\citeauthoryear{{Schaye} et~al.,}{{Schaye}
  et~al.}{2015}]{schaye15}
{Schaye} J.,  et~al., 2015, \mn@doi [\mnras] {10.1093/mnras/stu2058}, \href
  {http://adsabs.harvard.edu/abs/2015MNRAS.446..521S} {446, 521}

\bibitem[\protect\citeauthoryear{{Schiavon} et~al.,}{{Schiavon}
  et~al.}{2017}]{schiavon17}
{Schiavon} R.~P.,  et~al., 2017, \mn@doi [\mnras] {10.1093/mnras/stw3093},
  \href {https://ui.adsabs.harvard.edu/\#abs/2017MNRAS.466.1010S} {466, 1010}

\bibitem[\protect\citeauthoryear{{Trujillo-Gomez}, {Reina-Campos}  \&
  {Kruijssen}}{{Trujillo-Gomez} et~al.}{2019}]{trujillogomez19}
{Trujillo-Gomez} S.,  {Reina-Campos} M.,   {Kruijssen} J.~M.~D.,  2019, \mn@doi
  [\mnras] {10.1093/mnras/stz1932}, \href
  {https://ui.adsabs.harvard.edu/abs/2019MNRAS.488.3972T} {488, 3972}

\bibitem[\protect\citeauthoryear{{Usher}, {Pfeffer}, {Bastian}, {Kruijssen},
  {Crain}  \& {Reina-Campos}}{{Usher} et~al.}{2018}]{usher18}
{Usher} C.,  {Pfeffer} J.,  {Bastian} N.,  {Kruijssen} J.~M.~D.,  {Crain}
  R.~A.,   {Reina-Campos} M.,  2018, \mn@doi [\mnras] {10.1093/mnras/sty1895},
  \href {http://adsabs.harvard.edu/abs/2018MNRAS.480.3279U} {480, 3279}

\bibitem[\protect\citeauthoryear{{Webb} \& {Leigh}}{{Webb} \&
  {Leigh}}{2015}]{webb15}
{Webb} J.~J.,  {Leigh} N.~W.~C.,  2015, \mn@doi [\mnras]
  {10.1093/mnras/stv1780}, \href
  {http://adsabs.harvard.edu/abs/2015MNRAS.453.3278W} {453, 3278}

\bibitem[\protect\citeauthoryear{{Wiersma}, {Schaye}, {Theuns}, {Dalla Vecchia}
   \& {Tornatore}}{{Wiersma} et~al.}{2009}]{wiersma09b}
{Wiersma} R.~P.~C.,  {Schaye} J.,  {Theuns} T.,  {Dalla Vecchia} C.,
  {Tornatore} L.,  2009, \mn@doi [\mnras] {10.1111/j.1365-2966.2009.15331.x},
  \href {http://adsabs.harvard.edu/abs/2009MNRAS.399..574W} {399, 574}

\bibitem[\protect\citeauthoryear{{Zolotov}, {Willman}, {Brooks}, {Governato},
  {Brook}, {Hogg}, {Quinn}  \& {Stinson}}{{Zolotov} et~al.}{2009}]{zolotov09}
{Zolotov} A.,  {Willman} B.,  {Brooks} A.~M.,  {Governato} F.,  {Brook} C.~B.,
  {Hogg} D.~W.,  {Quinn} T.,   {Stinson} G.,  2009, \mn@doi [\apj]
  {10.1088/0004-637X/702/2/1058}, \href
  {http://adsabs.harvard.edu/abs/2009ApJ...702.1058Z} {702, 1058}

\makeatother
\end{thebibliography}

\bsp

\label{lastpage}

\end{document}